\documentclass[a4paper]{article}

\usepackage{INTERSPEECH2020}
\usepackage{multicol, blindtext}
\usepackage{color}
\usepackage{dsfont}
\usepackage{amsmath}
\usepackage{cite}
\usepackage{enumitem}
\usepackage{float}
\usepackage{array}
\usepackage{booktabs}
\newcolumntype{P}[1]{>{\centering\arraybackslash}p{\dimexpr#1\linewidth-2\tabcolsep}}

\title{Evaluating the reliability of acoustic speech embeddings}
\name{Robin Algayres$^{1,2}$, Mohamed Salah Zaiem$^{1,2}$, Benoît Sagot$^2$, Emmanuel Dupoux$^{1,2}$}
\address{
  $^1$ENS-PSL, EHESS, CNRS, Paris
  $^2$Inria, Paris
}
\email{robin.algayres@inria.fr, mohamed.zaiem@inria.fr, benoit.sagot@inria.fr, emmanuel.dupoux@inria.fr}

\begin{document}

\maketitle
\begin{abstract} 
Speech embeddings are fixed-size acoustic representations of variable-length speech sequences.
They are increasingly used for a variety of tasks ranging from information retrieval to unsupervised term discovery and speech segmentation. However, there is currently no clear methodology to compare or optimise the quality of these embeddings in a task-neutral way. Here, we systematically compare two popular metrics, ABX discrimination and Mean Average Precision (MAP), on 5 languages across 17 embedding methods, ranging from supervised to fully unsupervised, and using different loss functions (autoencoders, correspondence autoencoders, siamese). Then we use the ABX and MAP to predict performances on a new downstream task: the unsupervised estimation of the frequencies of speech segments in a given corpus. We find that overall, ABX and MAP correlate with one another and with frequency estimation. However, substantial discrepancies appear in the fine-grained distinctions across languages and/or embedding methods. This makes it unrealistic at present to propose a task-independent silver bullet method for computing the intrinsic quality of speech embeddings. There is a need for more detailed analysis of the metrics currently used to evaluate such embeddings.


\end{abstract}

\section{Introduction}

Unsupervised representation learning is the area of research that aims to extract units from unlabelled speech that are consistent with the phonemic transcription \cite{clsp,zs15,zs17}. As opposed to text, speech is subject to large variability. Two speech sequences with the same transcription can have significantly different raw speech signals. In order to work on speech sequences in an unsupervised way, there is a need for robust acoustic representations. To address that challenge, recent methods use {\em speech embeddings}, i.e.~fixed-size representations of variable-length speech sequences \cite{herman_cae,nils,settle,riad,emb2,emb3,emb5,cae}.\footnote{A speech sequence is a non-silent part of the speech signal (not necessarily a word). It can be transcribed into a phoneme $n$-gram.}

Speech embeddings can be used in many applications, such as key-word spotting\cite{query,query2,query3}, spoken term discovery\cite{utd,utd2,utd3}, and segmentation of speech into words \cite{goldwater,seg1,seg2}. It is convenient to evaluate the reliability of speech embeddings without being tied to a particular downstream task. One way to do that is to compute the intrinsic quality of speech embeddings. The basic idea is that a reliable speech embedding should maximise the information relevant to its type and minimise irrelevant token-specific information. Two popular metrics have been used: the mean average precision (MAP) \cite{map} and the ABX discrimination score \cite{abx}.

ABX and MAP are mathematically distinct yet they are expected to correlate well with each other as they both evaluate the discriminability of speech embeddings in terms of their transcription. However, \cite{nils} revealed a surprising result: the best model according to the ABX, is also the worst one according to the MAP. Following \cite{nils}'s results, we observed that this kind of discrepancies is much more common than we had expected. If a model performs well according to the MAP and bad according to the ABX, which metric should be trusted? For research in this field to go forward, there is a need to quantify the correlation of these two metrics.

In this paper, we wanted to go further and check that MAP and ABX can also predict performances on a downstream task. Such tasks are numerous, but one of them has not yet received enough interest: the \textit{unsupervised frequency estimation}. We define the frequency of a speech sequence as the number of times the phonetic transcription of this sequence appears in the corpus. When dealing with text corpora, frequencies can be computed exactly with a lookup table and are used in many NLP applications. In the absence of labels, deriving the frequency of a speech sequence becomes a problem of density estimation. Estimated frequencies can be useful in representation learning by enabling efficient sampling of tokens in a speech database \cite{riad}. Also, frequencies could be used for the unsupervised word segmentation using algorithms similar to those used in text \cite{goldwater}.

In Section~\ref{embeddings}, we present the range of embedding models that can be grouped in five categories of increasing expected reliability: hand-crafted, unsupervised, self-supervised, supervised plus a top-line embedding. In Section~\ref{tasks}, we present the MAP and ABX metrics and introduce our frequency estimation task. In Section~\ref{results}, we present results on the five speech datasets from the ZeroSpeech Challenge \cite{zs15,zs17}. From these results, we draw guidelines for future improvements in the field of acoustic speech embeddings.

\section{Embedding Methods}\label{embeddings}
\subsection{Acoustic features} 

Neural networks learn representations on top of input features. Therefore we used two types of acoustic features known as the log-MEL filterbanks (Mel-F)\cite{melf} and the Perceptual Linear Prediction (PLP)\cite{plp}. These two features can be considered as two levels of phonetic abstraction: a high-level one (PLP) and a low-level one (Mel-F).
Formally, let us define a speech sequence $s_t$ by $x_1$,$x_2$,...,$x_T$, where $x_i \in \mathbb{R}^n$ is called a frame of the acoustic features. $T$ is the number frames in the sequence $s_t$. In our setting, these frames are spaced out every 10 ms each representing a 25 ms span of the raw signal.

\subsection{Hand-crafted model: Gaussian downsampling}

Holzenberger and al. \cite{nils} described a method to create fixed-size embedding vectors that requires no training of neural networks: the Gaussian down-sampling (GD). Given a sequence $s_t$, $l$ equidistant frames are sampled and a Gaussian average is computed around each sample. It returns an embedding vector $e_t$ of size $l \times n$ for any size $T$ of input sequences. Therefore, given our two acoustic features, two baselines model are derived: the Gaussian-down-sampling-PLP (GD-PLP) and the Gaussian-down-sampling-Mel-F (GD-Mel-F).

Similarly, we derived a simple top-line model. Instead of using hand-crafted features, we can use the transcription of a given random segment. Each frame $x_i$ in a sequence $s_t$ will be assigned a 1-hot vector referring directly to the phoneme being said. This model goes through the same Gaussian averaging process to form the Gaussian-down-sampling-1hot (GD-1hot) model. This model is almost the true labels notwithstanding the information loss due to compression.

\subsection{Unsupervised model: RNNAE}

A more elaborate way to create speech embeddings is to learn them on top of acoustic features using neural networks. Specifically, recurrent neural networks (RNN) can be trained with back-propagation in an auto-encoding (AE) objective: the RNNAE \cite{nils,emb3}. Formally, the model is composed of an encoder network, a decoder network and a speaker encoder network. The encoder maps $s_t$ to $e_t$, a fixed-size vector. The speaker encoder maps the speaker identity to a fixed size vector $spk_t$. Then, the decoder concatenate $e_t$ and $spk_t$ and maps them to $\hat{s}_t$, a reconstruction of $s_t$. The three networks are trained jointly to minimise the  \textit{Mean Square Error} between $\hat{s}_t$ and $s_t$.

\subsection{Self-supervised and supervised models: CAE, Siamese and CAE-Siamese}

\subsubsection{Advanced training objectives}

We consider two popular embedding models. They are also encoder-decoders but they use additional side information. One is trained according to the Siamese objective \cite{riad,siamese,settle} the other is a correspondence auto-encoder (CAE) objective \cite{cae}. Both models assume a set of pairs of sequences from the training corpus. Positive pair are assumed to have the same transcription, negative pairs, different transcriptions. Let $p_t=(s_t,s_{t'},y)$ where $(s_t,s_{t'})$ is a pair of sequences  of lengths $T$ and $T'$. A binary value $y$ indicates the positive or negative nature of the pair. We will see how to find such pairs in the next sub-section. 

The CAE objective uses only positive pairs. The auto-encoder is asked to encode $s_t$ into $e_t$ and decode it into $\hat{s}_t$. The speaker encoder network is used similarly as for the RNNAE. To satisfy the CAE objective, $\hat{s}_t$ has to minimise the \textit{Mean Square Error} between $\hat{s}_t$ and $s_t'$. It forces the auto-encoder to learn a common representation for $s_t$ and $s_{t'}$ .

The Siamese objective does not need the decoder network. It encodes both $s_t$ and $s_{t'}$ and forces the encoder to learn a similar or different representation depending on whether the pair is positive or negative.
$$L_s(e_{t},e_{t'},y)=y \cos(e_t,e_{t'})\, -\, (1-y) \max(0,\cos(e_t,e_{t'})-\gamma)$$
where $\cos$ is the cosine similarity and $\gamma$ is a margin. This latter accounts for negative pairs whose transcriptions have phonemes in common. These pairs should not have embeddings 'too' far away from each other. The
CAE and Siamese objective can also be combined into a CAE-Siamese loss by a weighted sum of their respective loss function \cite{caesiamese}.

\subsubsection{Finding and choosing pairs of speech embeddings}

Finding positive pairs of speech sequences is an area of research called \textit{unsupervised term discovery} (UTD) \cite{utd,utd2,utd3,thual}. Such UTD systems can be DTW alignment based \cite{utd} or involve a k-Nearest-Neighbours search \cite{thual}. We opted for the latter, as it is both scalable and among the state-of-the-art methods. It encodes exhaustively all possible speech sequences with an embedding model, and used optimised $k$-NN search \cite{faiss} to retrieve acoustically similar pairs of speech sequences (see the details in \cite{thual}). In our experiments, we used the pairs retrieved by $k$-NN on \textit{GD-PLP} encoded sequences to train our self-supervised models (CAE,Siamese, CAE-Siamese).

As a supervised alternative, it is possible to sample `gold' pairs, i.e.~pairs of elements that have the exact same transcription. These `gold' pairs are given to the CAE, Siamese and CAE-Siamese to train supervised models. These supervised models indicate how good these self-supervised models could be if we enhanced the UTD system.

\section{Evaluation metrics and frequency estimation}\label{tasks}
\subsection{Intrinsic quality metrics: ABX and MAP}

The intrinsic quality of an acoustic speech embedding can be measured using two types of discrimination tasks: the MAP (also called same-different)\cite{map} and ABX tasks \cite{abx}.
Let us consider a set of $n$ acoustic speech embeddings: $((e_1,t_1),(e_2,t_2),...(e_n,t_n))$  where $e_i$ are the embeddings and $t_i$ the transcriptions. The ABX task creates all possible triplets ($e_a$,$e_b$,$e_x$) such that: $t_a = t_x$ and $t_b \neq t_x$. The model is asked to predict 1 or 0 to indicate if $e_x$ is of type $t_a$ or $t_b$. Such triplets are instances of the phonetic contrast between $t_a$ and $t_b$. Formally for a given a triplet, the task is to predict:
$$y(e_x,e_a,e_b)=\mathds{1}_{d(e_a,e_x) \leq d(e_b,e_x)}$$
The error rate on this classification task is the ABX score. It is first averaged by type of contrast (all triplets having the same $t_a$ and $t_b$) then average over all contrasts.\\
The MAP task forms a list of all possible pairs of embeddings ($e_a$,$e_x$). The model is asked to predict 1 or 0 to indicate if $e_x$ and $e_a$ have the same type, i.e the same transcription, or not. Formally for a given pair, the model predicts:
$$y(e_a,e_x,\theta) =\mathds{1}_{d(e_a,e_x) \leq \theta}$$
The precision and recall on this classification task are computed for various values of $\theta$. The final score or the MAP task is obtained by integrating over the precision-recall curve.

\subsection{Downstream task: unsupervised frequency estimation}
\subsubsection{The R$^2$ metric}
Here, we introduce the novel task of frequency estimation as the assignment, for each speech sequence, of a positive real value that correlate with how frequent the transcription of this sequence is in a given reference corpus\footnote{This estimation could be up to a scaling coefficient; the task of finding exact count estimates is a harder task, not tackled in this paper.}. To evaluate the quality of frequency estimates, we  use the correlation determinant $R^2$ between estimation and true frequencies. We compute this number in log space, to take into account the power-law distribution of frequencies in natural languages \cite{zipf}. This coefficient is between 0 and 1 and tells what percentage of variance in the true frequencies can be explained by the estimated frequencies. 

\subsubsection{$k$-NN and density estimation}
We propose to estimate frequencies using density estimation, also called the Parzen-Rosenblatt window method \cite{parzen}. Let $N$ be the number of speech sequence embeddings. First, these $N$ embeddings are indexed into a $k$-NN graph, noted $G$, where all distances between embeddings are computed. Then, for each embedding, we search for the $k$ closest embeddings in $G$. Formally, given an embedding $e_t$ from the $k$-NN graph $G$, we compute its $k$ distances to its $k$ closest neighbours ($d_{n_1}$,...$d_{n_k}$). The frequency estimation is a density estimation function $\kappa$ of the $k$-NN graph $G$ that has three parameter: a Gaussian kernel $\beta$, the number of neighbours $k$ and the embedding $e_t$.
$$\kappa_G(e_t,\beta,k)= \sum_{i=1}^{k} \exp^{-\beta d_{n_i}^2}$$

This density estimation yields a real number in $[1,k]$, which we take as our frequency estimation. We set $k$ to $2000$, the maximal frequency that should be predicted using the transcription of our training corpus (the Buckeye, see section 4.1). Then, we must tune $\beta$, the dilation of the space of a given embedding model. For each model, we choose $\beta$ such that it maximises the variance of the estimated log frequencies, thereby covering the whole spectrum of possible log frequencies, in our case $[0,\log(k)]$, which is beneficial for power-law types of distribution. Note that the $\beta$ kernel cannot be too large (resp.~small), as it would predict only high (resp.
~low) values.
\subsubsection{Density estimation versus clustering}
\begin{table}[H]
\centering\small
\begin{tabular}{lrrr}
  \toprule
   Models/methods & K-means & HC-K-means & $k$-NN \\
  \midrule
  GD-1hot & 0.67 & 0.73 & \textbf{0.74} \\
  RNNAE Mel-F & 0.30 & 0.35 & \textbf{0.41} \\
  CAE Siamese Mel-F & 0.26 & 0.37 & \textbf{0.43}  \\
  \bottomrule
\end{tabular}
\newline
\newline
\caption{Frequency estimations using K-means, HC-K-means and $k$-NN density estimation on a subset of the Buckeye corpus}\label{3}
\vspace{-2em}
\end{table}
We compared density estimation with an alternative method: the clustering of speech embeddings. Jansen et al.\cite{jansen} did a thorough benchmark of clustering methods on the task of clustering speech embeddings. Across all their metrics, the model that performs best is Hierarchical-K-means (HC-K-means), an improved version of K-means for a higher computational cost. In particular HC-K-means performs better than GMM HC-K-means is not scalable to our data sets, so we extracted 1\% of the Buckeye corpus in order to compare it with our method. A similar size of corpus is used by Jansen et al.\cite{jansen}.

We applied $k$-NN, K-means and HC-K-means from the python library scikit-learn \cite{scikit-learn} on three of our models on this subset. For K-means and HC-K-means, we used the hyper-parameters that gave the best scores in \cite{jansen}, namely k-means++ initialisation and average linkage function for HC-K-means. On our subset, the ground truth number of clusters is $K=33000$. Yet, we did a grid-search on the value of $k$ that maximises the $R^2$ score for frequency estimation. We found that K-means and HC-K-means perform better for $K=20000$. It shows these algorithms are not tuned to handle data distributed according to the Zipf's law. Indeed K-means is subject the so-called `uniform effect' and tends to find clusters of uniform sizes \cite{uniform}. Table \ref{3} shows that even by optimising the number of clusters $K$, the $k$-NN method outperforms K-means and HC-K-means. 

\section{Experiments}\label{results}
\subsection{Data sets}

Five data sets at our disposal from the ZeroSpeech challenge: Conversational English (a sample of the Buckeye \cite{pitt2005} corpus), English, French, Xitsonga and Mandarin \cite{zs15,zs17}. These are multi-speaker non-overlapping (i.e one speaker per file) recordings of speech. All silences were removed using \textit{voice activity detection} and corrected manually.

Each corpus was split into all possible segmentations to produce random speech sequences as described in \cite{thual}. Random speech sequences span from $70ms$ to $1s$. Shorter than $70ms$ sequences may contain less than one phoneme or be ill-pronounced phonemes. Therefore we removed very short sequences to avoid issues that are out of scope of this study.

The Buckeye sample corpus contains 12 speakers and 5 hours of speech. The French and English corpora being much larger, we reduced their number of speech sequences and speakers to the size of the Buckeye. Mandarin and Xitsonga are smaller data sets and were left untouched.
\begin{figure*}[t!]
  \centering
  \includegraphics[width=\linewidth]{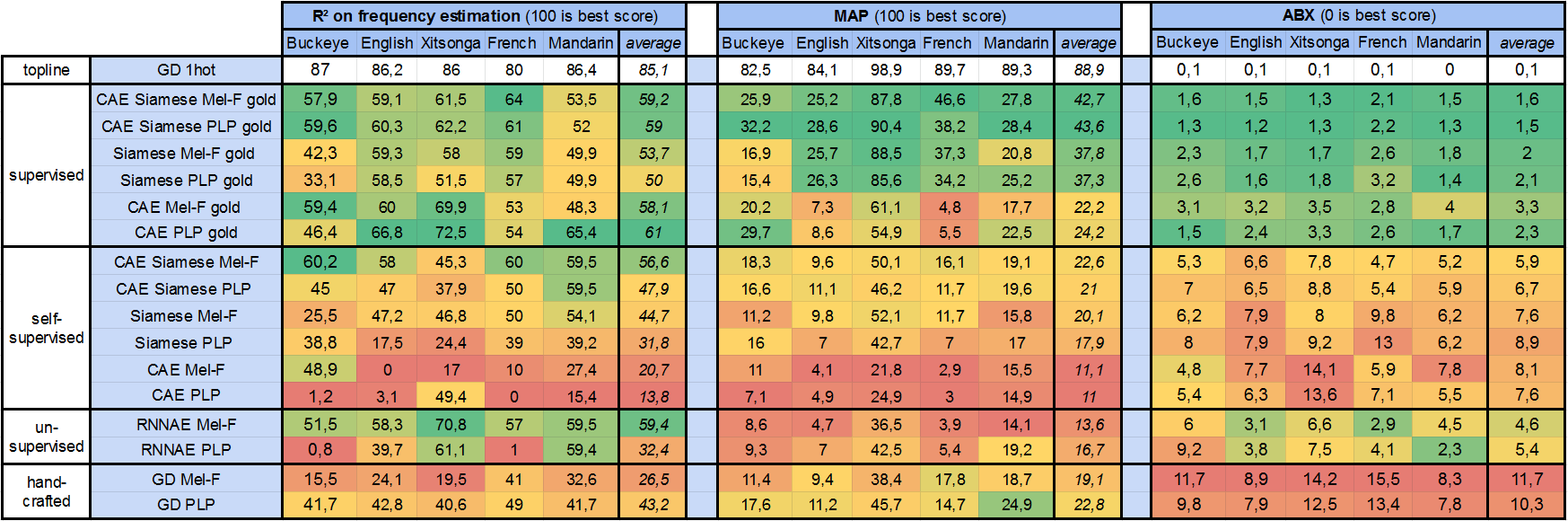}
  \caption{Value of metrics and the downstream task across models, corpora. The average column is the average score over all corpora}
  \label{1}
\end{figure*}

\subsection{Training and hyperparameters}
Our encoder-decoder network is a specific use of a three-layers bi-directional LSTM as described by Holzenberger et al. \cite{nils} with hyper-parameters selected to minimise the ABX error on the Buckeye corpus. The speaker embedding network is a single fully connected layer with fifteen neurons.
Our UTD system \cite{thual} uses the embeddings of the GD-PLP model. A set of speech pairs is returned, sorted by cosine similarity. We selected the pairs that have a cosine similarity above $0.85$ as it seemed to be optimal on the Buckeye corpus according to the ABX metric. In comparison, we trained our supervised models with `gold' pairs, i.e pairs with the exact same transcription. 

Each corpus was randomly split into train (90\%), dev(5\%) and test (5\%). Neural networks were trained on the train set, early stopping was done using the development set and metrics computed on the test set.  Specifically, we trained each model on the five training sets using the Buckeye's hyper-parameters. 

MAP was ABX were computed on the test sets. Frequency estimation was computed by indexing the five training sets and building $k$-NN graphs. For each element of a given test set, we searched neighbours and estimated frequencies using the $k$-NN graphs. We used the FAISS \cite{faiss} library that provides an optimised $k$-NN implementation.

\subsection{Results}

\subsubsection{Across models}

The results of the two metrics and downstream task are shown in Figure \ref{1} and the following broad trends can be observed.

\begin{itemize}[leftmargin=*]
\item Supervised models yield substantially lower performance than the ground truth 1-hot encodings, on all metric and all languages. These supervised models have a margin for improvement as they do not learn optimal embeddings despite having access to ground truth labels.

\item Supervised models outperform their corresponding self-supervised model, in almost all metrics and for all languages. It means that self-supervision has also a margin for improvement given better pairs from the UTD systems.

\item Among self-supervised and supervised models, the CAE-Siamese Mel-F takes the pole position. This model seems to be able to combine the advantages of both training objectives. A result already claimed by \cite{caesiamese}.

\item (self) supervised neural neural networks trained on low-level acoustic features (Mel-F) performs better or equally well as high-level acoustic features (PLP). This shows that neural networks can learn their own high-level acoustic features from low-level information.

\item Self-supervised models are expected to outperform unsupervised models because they use side information. Yet many configurations do not show this consistently. Only the Buckeye data set seems consistent, but this dataset is the one on which pairs were selected through a grid-search to minimise the ABX error. This may be due to the variable quality of the pairs found by UTD; better UTD is therefore needed to help self-supervised models.

\item Unsupervised models are supposed to be better than hand-crafted models because they can adjust by learning from the dataset. Yet, this is not consistently found.  Hand crafted models are worse than unsupervised models for ABX and frequency estimation but not for MAP.

\item In detail, which model is best in a particular language depends on the metric.
\end{itemize}

\subsubsection{Across metrics and frequency estimation}

In Table \ref{2}, we quantified the possibility to observe the discrepancies that we have just discussed . We computed the correlation $R^2$ across the three `average' columns. Cross correlation scores range from $R^2=0.33$ to $0.53$; the top-line model is not included when computing these scores. 
\begin{table}[H]
\centering\small
\begin{tabular}{lrrr}
  \toprule
    R$^2$ & Frequency est. & MAP & ABX \\
  \midrule
    Frequency est. & 1.0 &  0.34 & 0.53 \\
   MAP & 0.34 & 1.0 & 0.45 \\
   ABX & 0.53 & 0.45 & 1.0  \\
  \bottomrule
\end{tabular}
\newline
\newline
\caption{Correlation $R^2$ across the 'average' column of MAP, ABX and frequency estimation}\label{2}
\vspace{-2em}
\end{table}
These correlations are low enough to permit sizeable discrepancies across metrics and the downstream task. One of our model, the RNNAE Mel-F, epitomises the problem. This model is comparatively bad according to the MAP but good according to ABX and the frequency estimation. It means that MAP and ABX reveal different aspect of the reliability of embedding models. Therefore, only large progress according to one metric assures a progress according to an other metric. It shows the limit of intrinsic evaluation of speech embeddings. Moderate variations on a intrinsic metric cannot guarantee a progress on a given downstream task.

ABX and MAP scores are averages over multiple phonetic contrasts. These contrasts could be clustered based on their phonetic frequencies, average lengths or number of phonemes in common. Such fined-grained analyses can sometimes give understanding divergences across metrics. However, we have been unable to find a categorisation of results that make sense of Figure~\ref{1} as a whole. There are currently no fully reliable metrics to assess the intrinsic quality of speech embeddings.


\section{Conclusion}
We quantified the correlation across two intrinsic metrics (MAP and ABX) and a novel downstream task: frequency estimation. Although MAP and ABX agree on general categories (like supervised versus unsupervised embeddings), we also found large discrepancies when it comes to select a particular model highlighting the limits of these intrinsic quality metrics. However convenient intrinsic metrics may be, they only show partial views of the overall reliability of a model. We showed using frequency estimation that variations on intrinsic quality metrics should not be accounted for certain progress on downstream tasks. More attention should be brought on downstream tasks that have the credit to answer practical problems. \\
\vspace{-1em}
\section{Acknowledgements}
We thank Matthijs Douze for useful comments on density estimation. We also thank IDRIS from CNRS for offering GPU resources on the supercomputer Jean Zay. This work was funded in part by the Agence Nationale pour la Recherche (ANR-17-EURE-0017 Frontcog, ANR-10-IDEX-0001-02 PSL*, ANR-19-P3IA-0001 PRAIRIE 3IA Institute), CIFAR, and a research gift by Facebook.

\bibliographystyle{IEEEtran}

\bibliography{template}


\end{document}